\documentclass[twocolumn,english,10pt,journal,letterpaper]{IEEEtran}
\usepackage[T1]{fontenc}
\usepackage{color}
\usepackage{babel}
\usepackage{array}
\usepackage{float}
\usepackage{multirow}
\usepackage{amsmath}
\usepackage{amsfonts}
\usepackage{graphicx}
\usepackage{subcaption}
\usepackage{makecell}
\usepackage{cite}
\usepackage[none]{hyphenat}
\usepackage{hyperref}
\usepackage{gensymb}
\newcommand{\R}{\mathbb{R}}
\newcommand{\N}{\mathbb{N}}

\makeatletter

\renewcommand\footnoterule{ % Adding a 5cm line in the footnote
  \kern-3\p@
  \hrule\@width 5cm
  \kern2.6\p@}
  
\def\subsection{\@startsection{subsection}{2}{2.5\parindent}{1.5ex plus 1.5ex minus 0.5ex}%
{0.7ex plus .5ex minus 0ex}{\normalfont\normalsize\itshape}}%  

\makeatother

\begin{document}

% Specific conference options
\renewcommand{\figurename}{Fig.}
\renewcommand{\tablename}{TABLE}

\title{Simulation of a closed-loop dc-dc converter using a physics-informed neural network-based model}

\author{Marc-Antoine Coulombe, Maxime Berger, and Antoine Lesage-Landry 
% Est-ce que l'ordre a ici une importance outre le premier auteur? J'ai mis temporairement en ordre alphabétique de noms, mais la contribution me semble équitable pour vous deux.
% M.B. L'ordre est important, mais on peut le changer à la soumission finale si le papier est accepté.

\thanks{
M.-A. Coulombe and M. Berger are with the Department of Mathematics, Informatics, and Engineering, Université du Québec à Rimouski, Rimouski, Québec, Canada (e-mail of corresponding authors: [marc-antoine.coulombe, maxime\_berger]@uqar.ca).

A. Lesage-Landry is with the Department of Electrical Engineering, Polytechnique Montréal, GERAD \& Mila, Montréal, Québec, Canada (e-mail: antoine.lesage-landry@polymtl.ca). 

This work was funded by Mitacs, the Natural Sciences and Engineering Research Council of Canada (NSERC), and OPAL-RT Technologies under Grants ALLRP593314 \& IT40898.}

\thanks{Paper submitted to the International Conference on Power Systems Transients (IPST2025) in Guadalajara, Mexico, June 8-12, 2025.}

}

\maketitle
\begin{abstract}
The growing reliance on power electronics introduces new challenges requiring detailed time-domain analyses with fast and accurate circuit simulation tools. Currently, commercial time-domain simulation software are mainly relying on physics-based methods to simulate power electronics. Recent work showed that data-driven and physics-informed learning methods can increase simulation speed with limited compromise on accuracy, but many challenges remain before deployment in commercial tools can be possible. In this paper, we propose a physics-informed bidirectional long-short term memory neural network (BiLSTM-PINN) model to simulate the time-domain response of a closed-loop dc-dc boost converter for various operating points, parameters, and perturbations. A physics-informed fully-connected neural network (FCNN) and a BiLSTM are also trained to establish a comparison. The three methods are then compared using step-response tests to assess their performance and limitations in terms of accuracy. The results show that the BiLSTM-PINN and BiLSTM models outperform the FCNN model by more than 9 and 4.5 times, respectively, in terms of median RMSE. Their standard deviation values are more than 2.6 and 1.7 smaller than the FCNN's, making them also more consistent. Those results illustrate that the proposed BiLSTM-PINN is a potential alternative to other physics-based or data-driven methods for power electronics simulations.
\end{abstract}


\section{Introduction}\label{sec:Introduction}

\IEEEPARstart{T}{he} shift toward renewable sources of energy has accelerated, bringing a growing reliance on power electronics converters in power systems~\cite{Peyghami2021}. Their presence increases power system complexity and introduces new challenges related to stability, reliability, and safety. This requires the use of detailed time-domain analyses utilizing offline and real-time simulation tools to predict their impacts on power systems~\cite{Peyghami2020}. Fast and accurate simulation of power electronics remains challenging due to their time-varying structure, nonlinear response, and switching dynamics~\cite{Martinez2014}.  Moreover, recent applications require the use of real-time models and simulations, such as model predictive control and digital twins, further increasing the demand for faster circuit simulation~\cite{Vazquez2014, Tao2019, Wunderlich2021}. Commercially available electromagnetic transient (EMT) simulation software predominantly relies on physics-based methods to translate power electronics converters into equations that can be solved with classical numerical techniques~\cite{Berger2019}. Computation time reduction is generally done by substituting a time-varying with an equivalent fixed topology using switching functions~\cite{Salazar1994} or by using averaging techniques~\cite{Middlebrook1976, Sanders1991}. Recent works have shown that data-driven methods, physics-informed learning-based methods, and optimization techniques, can significantly increase the simulation speed while providing high-accuracy predictions, thus complementing traditional computational techniques~\cite{Ge2023, Rojas2020}. Data-driven and physics-informed learning-based methods are also valued for their ability to model commercial power electronics converters in a blackbox manner, removing the need for prior knowledge of the system~\cite{Qasqai2020, Rojas2020-blackbox}. These emerging approaches are still in their early stage, such that their performance requires deeper evaluation to make them suitable for reliable integration into commercial simulation tools.

Among the different machine learning-based modelling approaches used to model power electronics converters, neural network dominates, with primarily two architectures: feedforward neural networks (FNNs)~ \cite{Li2023, Krishnamoorthy2019, Zhao2022} and recurrent neural networks (RNNs)~\cite{Qasqai2020, Ge2023, Wunderlich2021}. In~\cite{Li2023}, a model with multiple fully-connected neural networks (FCNNs) is developed to model an IGBT's switching transients on a field-programmable gate array (FPGA) for real-time simulation. The authors achieved real-time simulation with a timestep of 5 ns. However, the proposed model requires 150 and 500 neural networks for simulating the turn-on and turn-off transient waveforms, respectively. In~\cite{Krishnamoorthy2019}, the authors combine an FCNN with Bayesian regularization backpropagation to model the transient response and random forests for the steady-state response. Using random forests alongside neural networks increases the model stability and reduces its variance, as the results obtained by decision trees are easier to interpret. An FCNN is used in~\cite{Zhao2022} to predict intermediate latent states between two observable states. Then, a physical model is used to predict the observable states using predictions from the intermediate latent states enabling them to integrate physics into their model. This approach requires solving multiple equations for each prediction using an implicit Runge-Kutta method, which can increase the computation time for complex models.
In~\cite{Qasqai2020}, a long-short term memory (LSTM) neural network is used to model a dc-dc buck converter. Although the LSTM appears to perform better than other FNN-based architectures for modelling the converter response, limited quantitative assements are provided to compare performance. Which is required for reliable deployment in simulation tools. In~\cite{Ge2023}, the output signal is decomposed into a transient and a periodic component modelled, respectively, by an FNN and an LSTM combined with a convolutional neural network. This approach reduces the complexity of the problem into two sub-problems. Reference~\cite{Wunderlich2021} uses a nonlinear autoregressive exogenous network, which yields good prediction accuracy. However, this type of architecture is limited in its ability to capture long-term dependencies without using internal memories like LSTMs. To consider a longer sequence of values, a large input size is required, which reduces computational efficiency.

While not directly applied to model power electronics converters, other data-driven approaches are also considered in the power systems literature, e.g., physics-informed neural ordinary differential equations for modelling power transformer's dynamic thermal behaviour \cite{Bragone2022} or for predicting the nonlinear transient dynamics of a inverter under fault \cite{Nellikkath2024} and graph-based learning for high-frequency filter design or microgrids voltage prediction \cite{LiY2023}. These approaches are promising alternatives for future studies on power electronics modelling.

Among the aforementioned articles on power electronics simulations, the following limitations have been observed:

\begin{itemize}
\item Some proposed solutions are difficult to generalize to multiple types of converters and different topologies;
\item The assessment of data-driven models is limited, i.e., with little to no quantitative metrics, thus, preventing their reliable deployment in simulation tools for power system analysis;
\item Some proposed solutions require significant computational time, making them difficult to apply in real-time simulations or without specialized hardware.
\end{itemize}

The main contributions of this paper are (i) designing a physics-informed bidirectional LSTM (BiLSTM-PINN) network architecture and to model the time-domain response of a closed-loop dc-dc boost converter and (ii) thoroughly benchmarking our model using step-response tests.
The BiLSTM architecture stands out as one well studied and documented, and easy to implement. It can also run on graphics processing units (GPUs) or FPGAs for computation acceleration \cite{Danopoulos2022, Nagarale2023}, making it readily accessible and practical for deployment in simulation software~\cite{Paszke2019, Rybalkin2018}. This architecture also takes into account both past and future values to make predictions, in contrast to RNNs or standard LSTMs, which only consider past values or FCNNs that are limited to present values. This property improves performance by utilizing the initial and final operating point values derived from the dc steady-state model to better reproduce the converter response. The inclusion of a physics-informed loss also helps mitigate prediction errors by penalizing results that do not adhere to physical principles during training.
Computational time evaluation is left for future work as many sources of interference such as the programming language (compiled MATLAB® \emph{vs.}~interpreted Python), the solver(s) used for the physical model or the hardware used, can make the results unreliable. This further underscores the importance of first defining the model architecture and assessing the model's accuracy, as it is done in this work. However, we consider that function evaluation with neural networks should be faster than solving a complex system of time-varying equations as hinted in~\cite{Ge2023}.

In this paper, we train an FCNN, a BiLSTM, and a BiLSTM-PINN to model the input and output currents of a closed-loop dc-dc boost converter's large-signal averaged model. The validation of the FCNN and BiLSTM models with a detailed switching model is beyond the scope of this paper. By using an averaged model, we aim to reduce the complexity of the neural network, enabling a more generalizable methodology and reduced computation time. Additionally, this model helps us quantify the transient waveforms using common performance metrics. This model is utilized to generate a dataset using MATLAB® Simulink for various operating points, parameters, and perturbations. The dataset is then used to train an FCNN, a BiLSTM and, a BiLSTM-PINN, with hyperparameter tuning performed via \texttt{HyperOpt}~\cite{Bergstra2013}. Finally, the two architectures are compared using step-response tests to assess model performance.

The paper is organized as follows. Section II presents the dc-dc boost converter topology and the large-signal averaged model used to generate the training and testing datasets. Section~\ref{sec:PINN} presents the physics-informed learning-based boost converter models. Section~\ref{sec:Numerical Experiments} discusses dataset generation and hyperparameter tuning, and validates the accuracy of our method via time-domain simulations in MATLAB® Simulink.

\section{Boost Converter Model}\label{sec:Boost}

\begin{figure}[H]
    \includegraphics[width=\columnwidth]{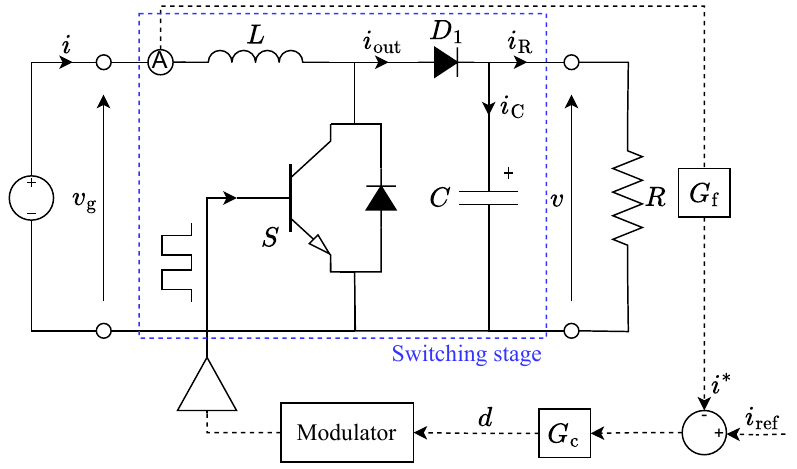}
    \caption{Circuit topology of the current regulated boost converter. \label{fig:boost}}
\end{figure}

The closed-loop boost converter topology used in this paper is shown in Fig.~\ref{fig:boost}. The switch $S$ is controlled by a pulse-width modulation (PWM) signal, introducing two distinct circuit states: switch~$S$ open and closed. The duty cycle $d(t)$ represents the proportion of time with the switch $S$ closed during a complete period $T_\text{s}$, while the complementary duty cycle is given by $d^\prime{} (t) = 1 - d(t)$. The input current $i$ is the regulated variable. The transfer function $G_\text{f}$ models the presence of a filter on the current measurement. %and $G_\text{c}$ is the controller which adjusts the duty cycle $d$ to track the reference current $i_\text{ref}$.

The circuit in Fig.~\ref{fig:boost} regulates the input current $i$ to match the reference $i_\text{ref}$ by calculating the error signal $\epsilon = i_\text{ref}-i^*$, where $i^*$ is the filtered input current. The error signal is then fed into a PI controller, with a transfer function defined as:
\begin{equation}\label{eq:controller_transfer_function}\notag
G_\text{c}(s)= \frac{sK_\text{p}+K_\text{i}}{s}\text{ ,}
\end{equation}
where $K_\text{p} \in \R$ and $K_\text{i} \in \R$ are determined using the \textit{algebra on the graph} technique \cite{Maksimovic2001} and further tuned to achieve the desired dynamic performance, with specified cross-over frequency $\omega_{\phi_\text{m}} = 2\pi f_{\phi_\text{m}}$ and phase-margin $\phi_\text{m}$ as explicited in Appendix. The filter and controller parameters for the converter model under validation are specified in Section~\ref{sec:Numerical Experiments}.

In dc steady-state, the dc conversion ratio $M(D)$, the inductor's dc current $I_\text{L}$, the inductor's current ripple $\Delta{i}_\text{L}$ and the capacitance's voltage ripple $\Delta{v}_\text{C}$ are derived as
\begin{align}
    M(D) & = \frac{V}{V_\text{g}} = \frac{1}{1-D}
    \label{eq:duty_cycle} \\
    I_\text{L} & = \frac{V}{(1-D)R}
    \label{eq:current_dc} \\
    \Delta{i}_\text{L} & = \frac{DV_\text{g}}{2f_\text{s}L}
    \label{eq:ripple_current_dc}\\
    \Delta{v}_\text{C} & = \frac{DV}{2f_\text{s}RC},
    \label{eq:ripple_voltage_dc}
\end{align}
where $V_\text{g}$ and $V$ are the dc input and output voltages, $D$ is the duty cycle, and $f_\text{s}$ is the switching frequency. We use ~\eqref{eq:duty_cycle}$-$\eqref{eq:ripple_voltage_dc} to describe the boost converter in initial and final steady-states which are used to initialize the BiLSTM models and the simulation parameters for the generation of the dataset in Section~\ref{sec:PINN} and Section~\ref{sec:Numerical Experiments}, respectively.

\begin{figure*}\centering
    \includegraphics[width=0.9\textwidth]{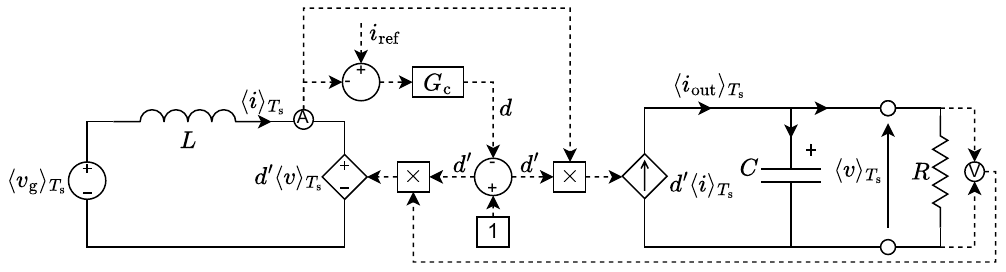}
    \caption{Closed-loop large-signal averaged model for the boost converter. \label{fig:boost_avm}}
\end{figure*}

We then use the classical dc-averaging technique for dc-dc converter~\cite{Erickson2020} to simplify the switching stage (in blue colour in Fig.~\ref{fig:boost}) leading to the following differential equations:
\begin{align}
    L\frac{\mathrm{d}\langle{i}\rangle_{T_\text{s}}}{\mathrm{d}t} & = \langle{v_\text{g}}\rangle_{T_\text{s}}-d^{\prime}\langle{v}\rangle_{T_\text{s}}
    \label{eq:voltage_avm} \\
    C\frac{\mathrm{d}\langle{v}\rangle_{T_\text{s}}}{\mathrm{d}t} & = d^{\prime}\langle{i}\rangle_{T_\text{s}}-\frac{\langle{v}\rangle_{T_\text{s}}}{R}, \label{eq:current_avm}
\end{align}
where denotes $\langle{}\rangle_{T_\text{s}}$ the average value of a variable over $T_\text{s}$. The equivalent circuit derived from~\eqref{eq:voltage_avm}~and~\eqref{eq:current_avm} is illustrated in Fig.~\ref{fig:boost_avm}. Our objective is to model the currents $\langle{i}\rangle{}_{T_\text{s}}$ and $\langle{i_\text{out}}\rangle{}_{T_\text{s}}$ in Fig.~\ref{fig:boost_avm} with learning-based methods. The generation of the dataset to train learning-based models using this circuit is presented in Section~\ref{sec:Numerical Experiments}.

\section{Physics-informed Learning-based Boost Converter Models}\label{sec:PINN}

We now develop our learning-based model of the boost converter. Let $\mathbf{x} \in \R^\text{18}$ and $\mathbf{y} \in \R^\text{2}$ be, respectively, the input (feature) and output (target) vector. We use the subscript $t \in \N$ to denote the time instance inputs/outputs are registered. We define the feature vector $\mathbf{x}_t$ as
% x = features, précisément
\begin{equation}\label{eq:features}
\mathbf{x}_t =  \textrm{vec}\begin{bmatrix}
t & L & C\\ R & R_\text{s} & V_\text{g} \\ V(t=0) & \Delta{v}(t=0) & I_\text{L}(t=0) \\ \Delta{i}_\text{L}(t=0) & I_\text{step} & t_\text{step} \\ K_\text{p} & K_\text{i} & f_\text{s} \\ f_\text{c} & i_{\text{ref},t} & d_{t}
\end{bmatrix},
\end{equation}
where \textrm{vec} is the vectorization operator. Various constants that characterize the converter are collected in \eqref{eq:features}, such as the inductance $L$, capacitance $C$, load resistance $R$, inductance resistance $R_\text{s}$, input voltage $V_\text{g}$, switching frequency $f_\text{s}$, and controller parameters $K_\text{i}$ and $K_\text{p}$, as derived in Section~\ref{sec:Boost}. Next, additional features are used to define the initial state of the system, namely, the output voltage $V(t=0)$, the capacitance's voltage ripple $\Delta{v}(t=0)$, the inductor's current $I_L(t=0)$, and the inductor's current ripple $\Delta{i}_\text{L}(t=0)$. Then, time-varying features, viz., the desired current $i_{\text{ref},t}$ and the control signal $d_{t}$ are included. Finally, $I_{\text{step}}$ and $t_{\text{step}}$ specify the amplitude and timing of the square disturbance on $i_{\text{ref},t}$.

In our setting, the entire system is assumed to be known, enabling the learning-based models to leverage comprehensive information about the circuit, its initial state, and its state at time $t$ for accurate predictions. This assumption is compatible with settings encountered in power system simulation tools. Further explorations could be done to reduce the dimension of the feature vector $\textbf{x}_t$ while minimizing the impact on accuracy. This is a topic for future work.

Next, we define the target vector at time $t$ as
\begin{equation}\notag
\mathbf{y}_t =  \begin{bmatrix}
\langle{i}\rangle{}_{T_\text{s},t} & \langle{i_\text{out}}\rangle{}_{T_\text{s},t}
\end{bmatrix}.
\end{equation}
The goal of the learning-based approach is to predict the input current $\langle{i}\rangle{}$ and the output current $\langle{i_\text{out}}\rangle{}$, derived using the previously defined large-signal averaged model. The decision to focus on current predictions is primarily to ensure signals operate on comparable time scales, hence simplifying the result analysis.

\begin{figure}[tb]
    \includegraphics[width=\columnwidth]{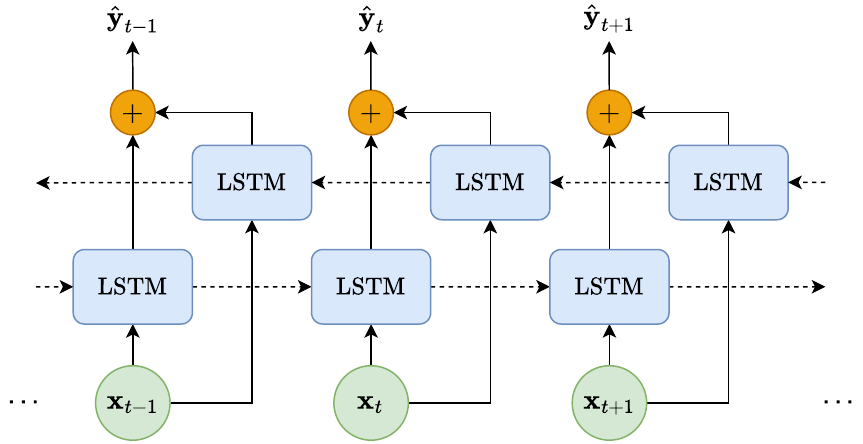}
    \caption{Schematic of a bidirectional LSTM neural network. \label{fig:bilstm}}
\end{figure}

Let $\texttt{FCNN}: \R^\text{18}\to \R^2$ and $\texttt{BiLSTM}: \R^\text{18}\to \R^2$ be, respectively, the trained FCNN and BiLSTM with their corresponding collections of weights denoted by $\mathcal{W}_\text{FCNN}$ and $\mathcal{W}_\text{BiLSTM}$. We first model the converter with an FCNN, a common architecture in the machine learning literature. The input vector~$\mathbf{x}_t$ is processed through the network, where a sequence of activation function and linear combination based weights $\mathbf{W} \in \mathcal{W}_\text{FCNN}$ compositions produces the prediction $\hat{\mathbf{y}}_{\text{FCNN}, t}$ at the output layer. The FCNN is later used for benchmarking. We then utilize a BiLSTM to model the converter dynamics. To generate a prediction $\hat{\textbf{y}}_{\text{BiLSTM,} t}$, the BiLSTM network processes features from $\textbf{x}_{t-k}$ to $\textbf{x}_t$ and $\textbf{x}_{t+k}$ to $\textbf{x}_{t}$ through LSTM cells, as illustrated in Fig.~\ref{fig:bilstm}. Here, $k \in \N$ denotes the sequence length, a hyperparameter that will be discussed in Section~\ref{sec:Numerical Experiments}. The weights $\mathbf{W} \in \mathcal{W}_\text{BiLSTM}$ are the trainable parameters within each LSTM cell fully-connected layers. Because the LSTM cell follows a well-documented topology, we refer interested readers to, e.g., \cite{VanHoudt2020, Qasqai2020}, for additional details. Finally, given observation $\mathbf{x}_t$, the resulting predictions $\mathbf{\hat{y}}_t$ are:
\begin{align}
    \hat{\textbf{y}}_{\text{FCNN}, \textit{t}} & = \texttt{FCNN}(\textbf{x}_{t}; \mathcal{W}_\text{FCNN}) \label{eq:model_FCNN} \\
    \hat{\textbf{y}}_{\text{BiLSTM}, \textit{t}} & = \texttt{BiLSTM}(\textbf{x}_{t}; \mathcal{W}_\text{BiLSTM}). \label{eq:model_BiLSTM}
\end{align}

To achieve accurate predictions, the collections of weights $\mathcal{W}_\text{FCNN}$ and $\mathcal{W}_\text{BiLSTM}$ in \eqref{eq:model_FCNN} and \eqref{eq:model_BiLSTM} are computed to minimize the discrepancy between the predictions $\hat{\textbf{y}}$ and the targets $\textbf{y}$ through a loss function. Let the loss function be defined as
\begin{equation}\label{eq:loss}
\mathcal{L} = \mathcal{L}_\text{RMSE} + \lambda \mathcal{L}_\text{PBE},
\end{equation}
where $\mathcal{L}$ combines the root-mean-squared error $\mathcal{L}_\text{RMSE}$ with the circuit power balance error $\mathcal{L}_\text{PBE}$ weighted by a constant $\lambda>0$ which is to be tuned. The loss function is used to compute the gradient, which drives the backpropagation algorithm that trains both networks. By iteratively adjusting the collections of weights $\mathcal{W}_\text{FCNN}$ and $\mathcal{W}_\text{BiLSTM}$ using \eqref{eq:loss}, the models progressively improve their accuracy. Specifically, the loss function $\mathcal{L}_\text{RMSE}$ for a single curve is
\begin{equation}\label{eq:loss_rmse}
\mathcal{L}_\text{RMSE} = \sqrt{\frac{1}{n}\sum_{t=1}^n(\hat{\textbf{y}}_t-\textbf{y}_t)^2} \text{ ,}
\end{equation}
where $n \in \N$ is the number of points on a curve related to the simulation duration and the step size. The power balance loss function $\mathcal{L}_\text{PBE}$ is then given by:
\begin{equation}\label{eq:loss_pbe}
\mathcal{L}_\text{PBE} = \frac{1}{n}\sum_{t=1}^n |\langle{P_\text{in}}\rangle{}_{T_\text{s},t}-\langle{P_\text{out}}\rangle{}_{T_\text{s},t}|,
\end{equation}
where the predicted currents $\hat{\textbf{y}}_t$ are used to calculate, respectively, the predicted power entering and exiting the converter, defined as:
\begin{align}
    \langle{P_\text{in}}\rangle{}_{T_\text{s},t} & = d^\prime{}_{t}\langle{v}\rangle{}_{T_\text{s},t} \langle{i}\rangle{}_{T_\text{s},t} \label{eq:loss_power_in} \\
    \langle{P_\text{out}}\rangle{}_{T_\text{s},t} & = \langle{v}\rangle{}_{T_\text{s},t}  \langle{i_\text{out}}\rangle{}_{T_\text{s},t} \text{ .} \label{eq:loss_power_out}
\end{align}
In other words, we consider that the conservation of power applies while neglecting the converter inner losses. This assumption allows us to integrate physics into the BiLSTM loss function by penalizing predictions yielding a high-power imbalance between the converter's input and output, then leading to the BiLSTM-PINN (and FCNN) model.

\section{Numerical Results}\label{sec:Numerical Experiments}
% In Section \ref{sec:Numerical Experiments}, two parts are explicited: (i) the numerical setting, how to reproduce the dataset and the training with tuned hyperparameters and (ii) the benchmarking of our proposed model.
We now present our numerical setting and then evaluate the performance of our learning-based models.
\subsection{Numerical setting and training}

To conduct the experiment, we first generate our dataset. This dataset is created by reproducing the lossless switching circuit in Fig.~\ref{fig:boost} and the large-signal averaged model shown in Fig.~\ref{fig:boost_avm} in MATLAB® Simulink. The simulation parameters to create the dataset are provided in Table~{\ref{tab:simulation parameters}}.

\begin{table}[htbp]
    \centering
    \caption{Simulations Paramaters for the Dataset Generation.}
    \label{tab:simulation parameters}%
    \begin{tabular}{ccccc}
\hline

\hline
\textbf{Parameter} & \textbf{Interval} & \textbf{Step size} \\ \hline
$C$ (\text{$\mu${}}F) & [1000, 7000] & 1000 \\
$d(t)$& $\pm 0.9$ & -- \\
$f_\text{c}$ (kHz) & 5 & -- \\ 
$f_\text{s}$ (kHz) & 70 & -- \\
$f_{\phi_\text{m}}$ (Hz) & 255 & -- \\
$I_\text{step}$ (A) & $\pm I_L(t=0)$ & -- \\
$L$ (mH) & [1, 7] & 1 \\
$R$ ($\Omega$) & [1, 500] & 10 \\
$R_\text{s}$ ($\Omega$) & 30$\times10^{-3}$ & -- \\
$V$ (V) & [200, 400] & 10 \\
$V_\text{g}$ (V) & [100, 200] & 10 \\
$t$ (s) & [0, 0.03] & 250$\times10^{-7}$ \\
$t_\text{step}$ (s) &  [0.011, 0.021] & -- \\
$\phi_\text{m}$ (\textdegree{}) & 50 & -- \\
\hline

\hline
\end{tabular}
\end{table}
Key parameters of Fig.~\ref{fig:boost_avm}'s circuit are defined, with their study range specified in a [lower bound, upper bound] format along with the step size for varying parameters such as $C$, $L$, $R$, $V$, and $V_\text{g}$. Other parameters, like $f_\text{s}$ and $R_\text{s}$, remain constant for all simulations. During a simulation, a current step of amplitude $I_\text{step}$ is applied to $i_\text{ref}$ at time $t_\text{step}$ based on a uniform distribution. The simulation itself is carried out from $t = 0$ to $t = 0.03$ with a fixed time step of $250 \times 10^{-7}$s. 

In total, 565,950 operating points are simulated. Among these, cases that do not exhibit continuous conduction mode are excluded because they display response not accounted for by the large-signal averaged model. We also removed cases with an unfeasible PI controller (see Appendix). In addition, any cases where $|d| = 0.9$ are also rejected because the controller is programmed to saturate at this value, and saturation introduces non-linearities outside of the scope of this study. Approximately 10\% of the dataset is removed for these reasons. Additionally, the points corresponding to the first 0.01 second of each simulation are discarded to ensure the model reaches steady-state before introducing a perturbation. Finally, the dataset is split randomly using the holdout method with an 80$\%-$20\% ratio.

We identify suitable hyperparameters for our learning-based models using \texttt{HyperOpt} \cite{Bergstra2013}. The resulting hyperparameters are presented in Table~{\ref{tab:hyperparameters}} together with the interval considered by \texttt{HyperOpt}. To define the intervals, preliminary trials were conducted to establish practical ranges, avoiding underfitting or overfitting. These issues were identified through an observable increase in the loss function $\mathcal{L}$. Subsequently, 30 optimization trials were performed using the tree-structured Parzen estimator (TPE) algorithm, with the objective of minimizing $\mathcal{L}$. The resulting values were rounded for simplicity. We chose the \texttt{ADAM} optimizer for all trials as it consistently provided better results compared to the alternatives \cite{Kingma2017}.
From the hyperparameters described in Table~\ref{tab:hyperparameters}, the FCNN, BiLSTM, and BiLSTM-PINN models are trained on 80\% of the dataset. Because our model is trained offline, we train multiple models simultaneously and select the one that delivers the best testing RMSE. Although this approach does not fully address sensitivity issues, it helps mitigate their impact by choosing the best-performing empirical model for deployment. Robustness of the model, the stability of outputs across different inputs, still remains to be established similarly to many of the neural network models. This is a topic for future work.

\begin{table}[htbp]
    \centering
    \caption{Hyperparameters interval and tuned values}
    \label{tab:hyperparameters}%
    \begin{tabular}{ccccc}
\hline

\hline
\textbf{Hyperparameter} & \textbf{Interval} & \texttt{\textbf{FCNN}} & \texttt{\textbf{BiLSTM(-PINN)}} \\ \hline
Batch size &  [128, 300] & 294 & 207 \\
Epochs & -- & 35 & 35 \\
Learning rate & [1$\mathrm{e}^{-4}$, 1$\mathrm{e}^{-2}$] & 6$\mathrm{e}^{-4}$ & 1$\mathrm{e}^{-3}$ \\ 
Optimizer & -- & \texttt{ADAM} & \texttt{ADAM} \\
\thead{Neurons/Cells \\ (per hidden layer)} & [100, 500] & 304 & 165 \\
Sequence length $k$ & [10, 25] & -- & 25 \\
Weights decay & [1$\mathrm{e}^{-5}$, 1$\mathrm{e}^{-2}$] & 9$e^{-3}$ & 7$\mathrm{e}^{-4}$ \\
$\lambda$ & [0, 1] & 0.2 & 0 (0.2) \\
\hline

\hline
\end{tabular}
\end{table}

\subsection{Test}

\begin{figure}[tb]

    \begin{subfigure}{\columnwidth}
        \includegraphics[width=1\textwidth,trim={0 1.5cm 0 0.5cm}]{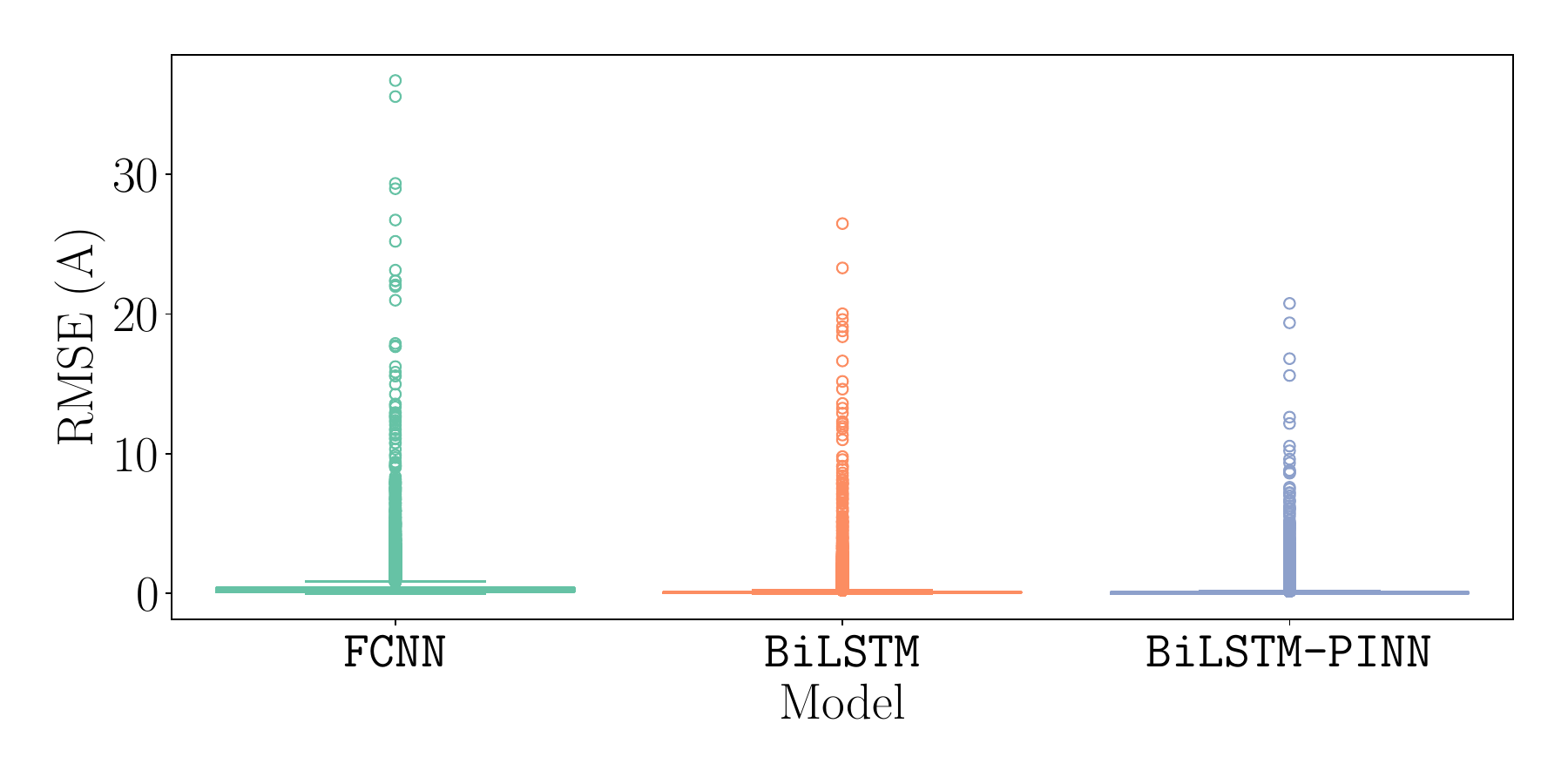}
        \caption{RMSE distribution at the FCNN scale. Circles represent the outliers of each model on the test dataset.}
        \label{fig:Boxplot}
    \end{subfigure}
    
    \begin{subfigure}{\columnwidth}
        \includegraphics[width=1\textwidth,trim={0 1.5cm 0 0.25cm}]{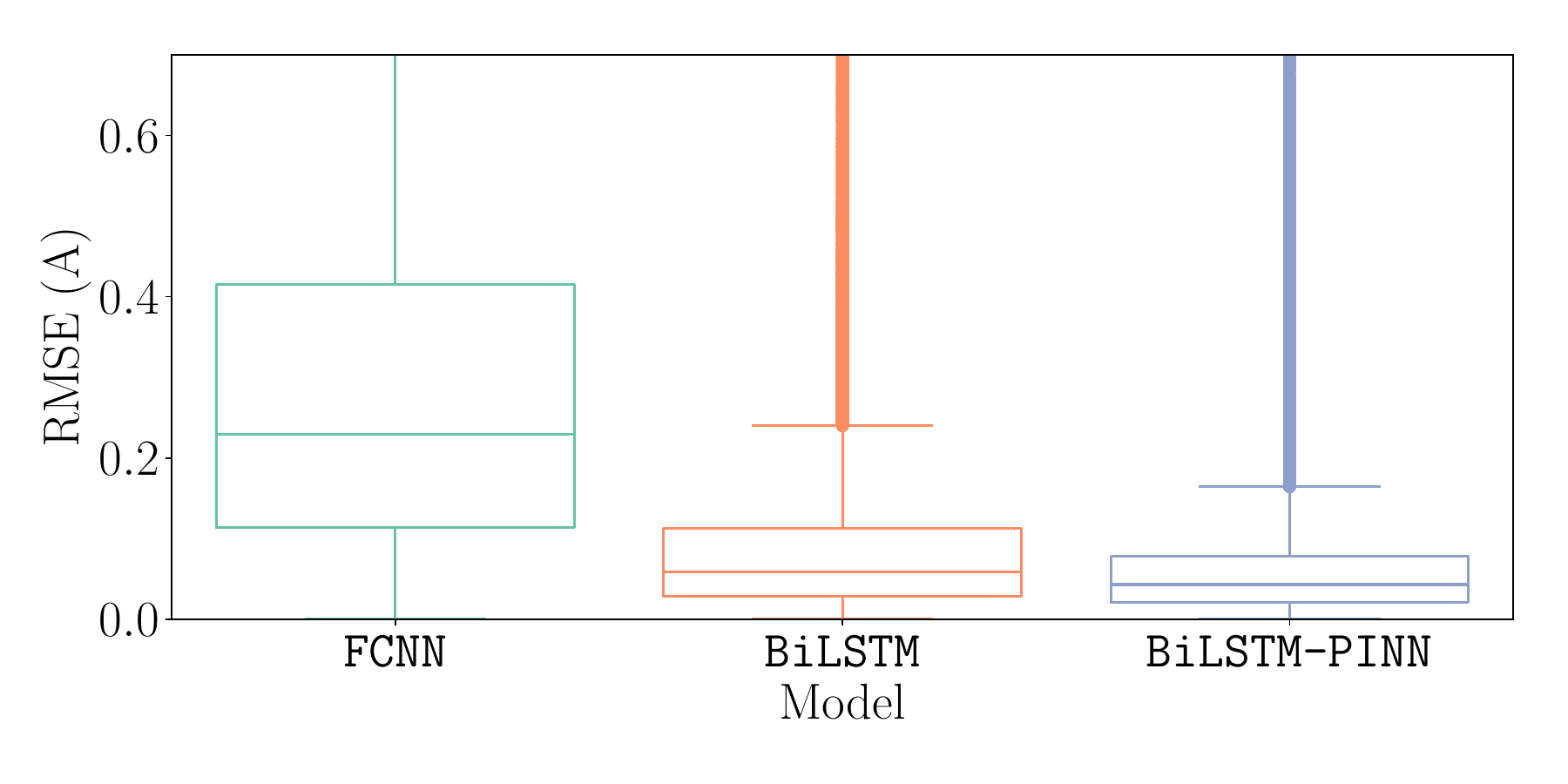}
        \caption{Quartiles comparison for both learning-based models representing quartiles and median values. The lowest and highest delimitations of the box are the $25^\text{th}$ and $75^\text{th}$ percentile, respectively. The vertical line inside the box represents the median and circles (appearing as a bold line) are the RMSE distribution outliers.}
        \label{fig:Boxplot_zoomed}
    \end{subfigure}
    \vspace{-0.5cm}
    \caption{RMSE box plot for the FCNN, BiLSTM, and BiLSTM-PINN models.}
    \label{fig:box_plots}
\end{figure}

We then use the remaining 20\% of the dataset for out-of-sample testing purposes. For each predicted response curves, we calculate the RMSE using~\eqref{eq:loss_rmse}. The resulting box plot is illustrated at Fig.~\ref{fig:box_plots}.

Fig.~\ref{fig:box_plots} presents the RMSE for the FCNN, BiLSTM, and BiLSTM-PINN models. From Fig.~\ref{fig:Boxplot}, we can observe that some operating point's RMSE greatly diverges from the median value. This is especially the case for the FCNN. Fig.~\ref{fig:Boxplot_zoomed} further zooms on the box plot distribution. We remark that the BiLSTM-PINN outperforms the FCNN and the BiLSTM by achieving a lower median $\mathcal{L_\text{RMSE}}$ and distribution dispersion.

Next, we quantified the model prediction performance using the \texttt{stepinfo} function from MATLAB®. We refer readers to their documentation for more details on the function and the evaluated metrics~\cite{MATLAB}. Table~\ref{tab:metrics} compares all performance metrics for the FCNN and BiLSTM models.

\begin{table*}[htbp]
    \centering
    \caption{Learning-based model performance metric comparison}
    \begin{tabular}{ccccc}
    
\hline

\hline
\textbf{Metric} & \texttt{\textbf{FCNN}} & \texttt{\textbf{BiLSTM}} & \texttt{\textbf{BiLSTM-PINN}} \\
& $\text{mean} (e) \pm \sigma_\text{std} (e)$ & $\text{mean} (e) \pm \sigma_\text{std} (e)$ & $\text{mean} (e) \pm \sigma_\text{std} (e)$ \\ & $\text{median}(e)$ & $\text{median}(e)$ & $\text{median}(e)$ \\ \hline
$\mathcal{L_\text{RMSE}}$ (A) & \thead{0.3793 $\pm$ 0.7367 \\ 0.2311} & \thead{0.1161 $\pm$ 0.4236 \\ 0.0509} & \thead{0.0593 $\pm$ 0.2782 \\ 0.0256} \\
Rise time (s) & \thead{0.0004 $\pm$ 0.0009 \\ 0.0000e-5} & \thead{7.7094e-5 $\pm$ 0.0004 \\ 0.0000e-5}
& \thead{3.5621e-5 $\pm$ 0.0003 \\ 0.0000e-5}\\
Transient time (s) & \thead{0.0036 $\pm$ 0.0037 \\ 0.0022} & \thead{0.0034 $\pm$ 0.0032 \\ 0.0024}
& \thead{0.0022 $\pm$ 0.0021 \\ 0.0015}\\
Settling time (s) & \thead{0.0020 $\pm$ 0.0026 \\ 0.0010} & \thead{0.0016 $\pm$ 0.0023 \\ 0.0007} &
\thead{0.0019 $\pm$ 0.0031 \\ 0.0003}\\ 
Settling minimum (A) & \thead{0.3131 $\pm$ 0.9630 \\ 0.1892} & \thead{0.1205 $\pm$ 1.5329 \\ 0.0438} & 
\thead{0.0685 $\pm$ 0.9676 \\ 0.0281}\\
Settling maximum (A) & \thead{0.5400 $\pm$ 1.6805 \\ 0.2034} & \thead{0.0747 $\pm$ 0.3949 \\ 0.0277} & 
\thead{0.1131 $\pm$ 0.8193 \\ 0.0299}\\
Overshoot (\%) & \thead{{--} \\ 7.4707} & \thead{{--} \\ 1.8219} &
\thead{{--} \\ 1.7154}\\
Undershoot (\%) & \thead{{--} \\ 0.0000e-5} & \thead{{--} \\ 0.0000e-5} &
\thead{{--} \\ 0.0000e-5}\\
Peak (A) & \thead{0.5373 $\pm$ 1.6714 \\ 0.2034} & \thead{0.0741 $\pm$ 0.3933 \\ 0.0277} &
\thead{0.1100 $\pm$ 0.7640 \\ 0.0299}\\
Peak time (s) &\thead{0.0027 $\pm$ 0.0031 \\ 0.0008} & \thead{0.0017 $\pm$ 0.0029 \\ 0.0002} &
\thead{0.0029 $\pm$ 0.0037 \\ 0.0005} &\\
\hline

\hline
\label{tab:metrics}%
\end{tabular}
\end{table*}

\begin{figure}[h]
    \begin{subfigure}[b]{\columnwidth}
        \includegraphics[width=0.98\textwidth]{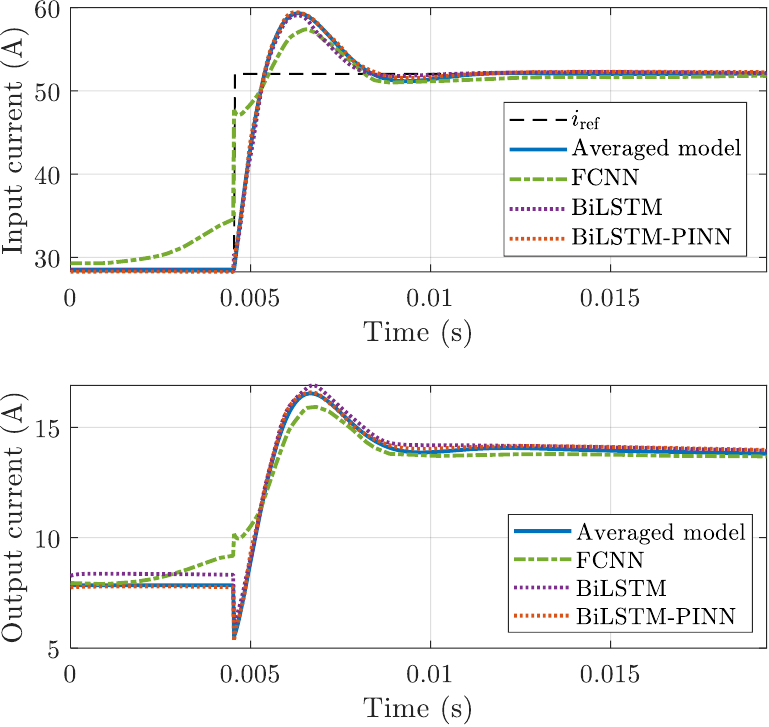}
        \caption{Input and output current signals for $\approx83\%$ increase in $i_\text{ref}$.}
        \label{fig:pred1}
    \end{subfigure}\medskip
    
    \begin{subfigure}[b]{\columnwidth}
        \includegraphics[width=0.98\textwidth]{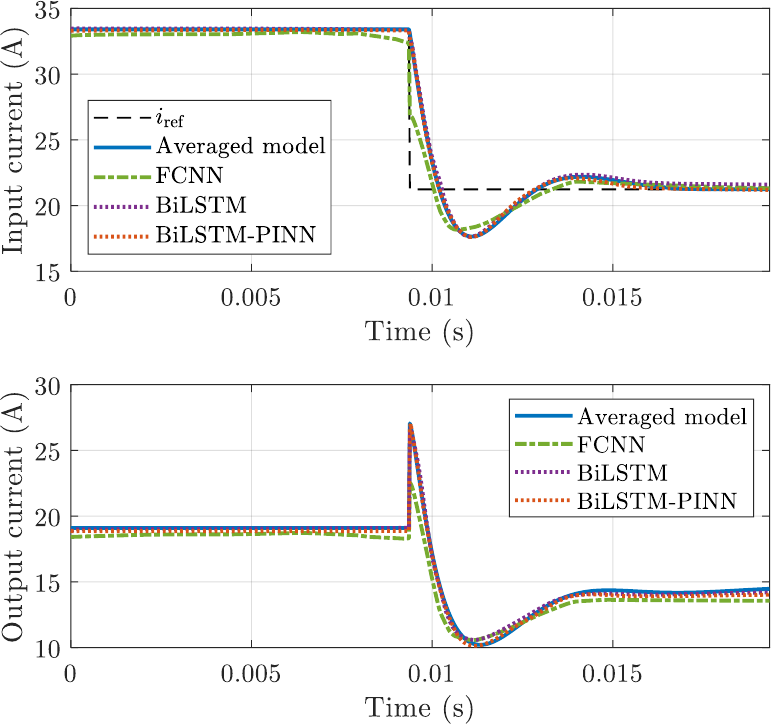}
        \caption{Input and output current signals for $\approx36\%$ decrease in $i_\text{ref}$.}
        \label{fig:pred2}
    \end{subfigure}\medskip

    \caption{Predictions for two operating points achieving close to or lower median performance in all metrics evaluated in Table~\ref{tab:metrics}.}
    \label{fig:pred1_2}
\end{figure}

\begin{figure}[h]
    \begin{subfigure}[b]{\columnwidth}
        \includegraphics[width=0.99\columnwidth]{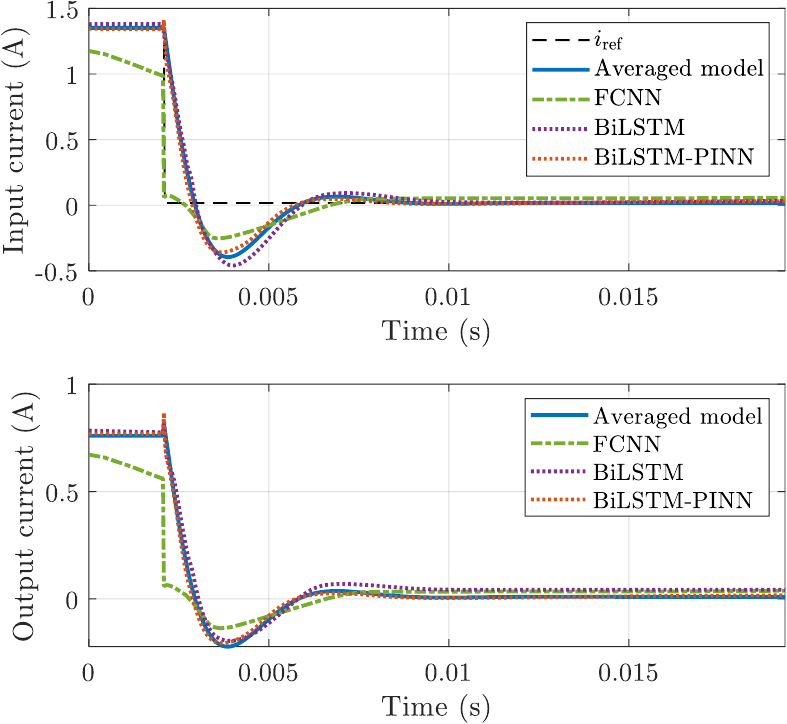}
        \caption{Input and output currents signals for $\approx99\%$ decrease on $i_\text{ref}$.}
        \label{fig:pred3}
    \end{subfigure}\medskip

% *********** compléter celle-ci *************
    \begin{subfigure}[b]{\columnwidth}
        \includegraphics[width=0.99\columnwidth]{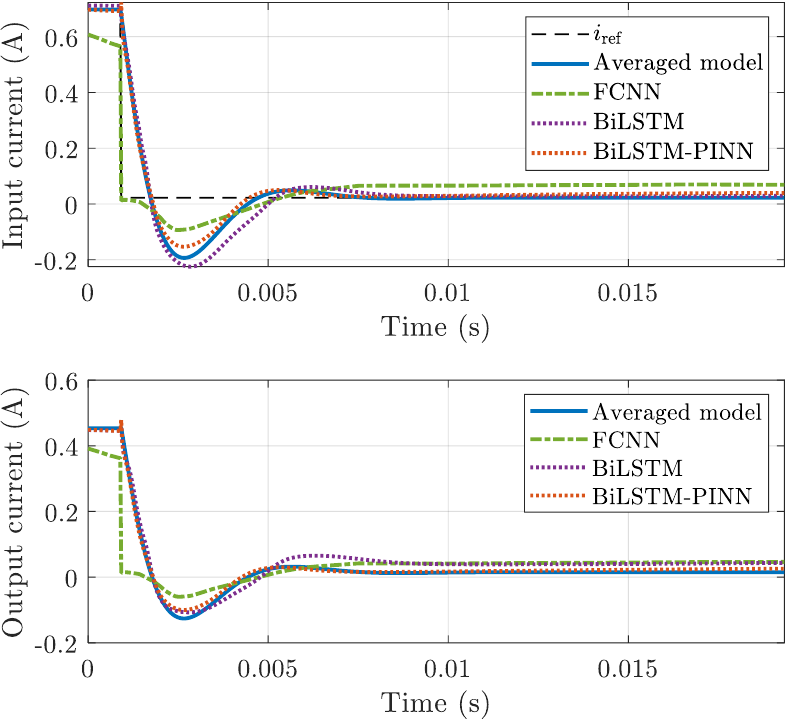}
        \caption{Input and output currents signals for $\approx98\%$ decrease on $i_\text{ref}$.}
        \label{fig:pred4}
    \end{subfigure}\medskip
% *********** compléter celle-ci *************    
    % \vspace{-0.5cm}
    
    \caption{Predictions for two operating points resulting in some of the highest overshooting error (a) or undershooting error (b) among all operating points tested in Table~\ref{tab:metrics} (outliers).}
    % \label{fig:preds}
    \label{fig:pred3_4}
\end{figure}

Recall that $\mathbf{y}_t$ ($\mathbf{\hat{y}}_t$) consists of the input (predicted) current $\langle{i}\rangle{}_{T_\text{s}, t}$ ($\langle{\hat{i}}\rangle{}_{T_\text{s}, t}$) and the (predicted) output $\langle{i_\text{out}}\rangle{}_{T_\text{s},t}$ ($\langle{\hat{i}}_\text{out}\rangle{}_{T_\text{s},t}$). Let the total absolute prediction error be
\begin{equation}\label{eq:error} \nonumber
e = e_\text{input}+e_\text{output},
\end{equation}
where $e_\text{input}$ and $e_\text{output}$ are defined as the absolute input and output error, respectively, given by
\begin{align*}
e_\text{input} &= \left|\langle{\hat{i}}\rangle{}_{T_\text{s},t}-\langle{i}\rangle{}_{T_\text{s},t}\right|\\
e_\text{output} &= \left|\langle{\hat{i}_\text{out}}\rangle{}_{T_\text{s},t}-\langle{i_\text{out}}\rangle{}_{T_\text{s},t}\right|.
\end{align*}
For each \texttt{stepinfo} characteristics, we report the mean, the standard deviation $\sigma_\text{std}$, and the median. Table~\ref{tab:metrics} compares all performance metrics for the FCNN, BiLSTM, and BiLSTM-PINN models. Table~{\ref{tab:metrics}} illustrates the superiority of BiLSTM-PINN over FCNN and BiLSTM on most metrics in terms of mean($e$) and median($e$). We remark that $\sigma_\text{std}(e)$ is often bigger than mean($e$), demonstrating that outliers are, skewing the mean, similarly to the RMSE analysis in Fig.~\ref{fig:box_plots}. In this case, the median might be more informative as an overall evaluation of the performance. Looking at the RMSE medians, we observe that the BiLSTM-PINN has a median error of 0.0256~A ($\approx9$ times lower than the FCNN) in comparison to 0.0509~A ($\approx4.5$ times lower than the FCNN), and 0.2311~A for the BiLSTM and the FCNN, respectively. We similarly notice that standard deviation values are 0.2782~A, 0.4236~A, and 0.7367~A for the BiLSTM-PINN, BiLSTM, and FCNN. The BiLSTM-PINN and BiLSTM standard deviation are approximately 2.6 and 1.7 times smaller than the FCNN, making the predictions not only generally more accurate, but also more consistent. Those results are consistent with \cite{Qasqai2020}, which showed that recurrent architecture tends to outperform non-recurrent ones. However, unlike \cite{Ge2023}, our physics-informed FCNN cannot reproduce as accurately the waveform when using $\mathcal{L}_\text{PBE}$ to integrate physics to the loss. Over all metrics, the BiLSTM-PINN main limitations are the overshoot and the undershoot. To better understand the problem, we extract four distinct cases from the test dataset, two with average performance and two with a high overshooting or undershooting error and plot their responses. Figs.~\ref{fig:pred1_2} and~\ref{fig:pred3_4}, respectively, represent the two cases with average performance and with a high overshooting or undershooting error. The BiLSTM-PINN tends to follow accurately the averaged model in both cases while the FCNN exhibits some inaccuracies, e.g., Fig.~\ref{fig:pred3_4}. We notice that overshooting or undershooting error occurs when there is a small perturbations on $i_\text{ref}$ and the final value is close to zero as observed in Fig.~\ref{fig:pred3_4}. These cases tend to produce outliers in terms of overshooting and undershooting errors given the division to a value close to zero, skewing the distribution and making the mean and standard deviation less representative. We omitted these values from Table~\ref{tab:metrics} for this reason, keeping only the median values as more insightful in this case. This phenomenon is also amplified by the minimum and maximum values on $I_\text{L}(t=0)$ in our dataset. For perturbations on $i_\text{ref}$ close to $|I_\text{L}(t=0)|$, a reduced number of operating points exist for training in our dataset due to the use of a uniform distribution, thus increasing the error of our model in these cases. In sum, the results illustrate that our models may achieve good performance inside the parameter intervals defined in Table~\ref{tab:hyperparameters}, but lose accuracy when operating closer to their minimum and maximum values due to having fewer available operating training points.

% \vspace{0.5cm}
%\begin{figure}[H]
%    \includegraphics[width=0.97\columnwidth]{Figures/pred3.pdf}
%    \caption{Input and output currents signals for an operating point with a high %overshooting error. \label{fig:pred3}}
%\end{figure}

\section{Conclusions}\label{sec:Conclusion}

This work proposes a new physics-informed BiLSTM method to model power electronics converters. We evaluate the proposed BiLSTM-PINN by generating a dataset using the steady-state equations of a dc-dc boost converter along with a dc-averaged closed-loop model of the converter in MATLAB® Simulink. We also train a physics-informed FCNN and a data-driven BiLSTM to establish a performance comparison with a conventional non-recurrent neural network architecture. For both architectures, we tune the hyperparameters using \texttt{HyperOpt}. We then compare both methods by visualizing the RMSE distribution and quantifying their accuracy for various metrics using the \texttt{stepinfo} function from MATLAB®.

Our numerical study illustrates the ability of learning-based models to predict the response of power electronics converters. The BiLSTM-PINN model offers improved performance over the FCNN's and the data-driven BiLSTM and may be a good alternative to classical methods in some applications as a means of simplifying and accelerating real-time simulations. These results can improve the capabilities of commercial EMT simulation software by allowing the simulation of large-scale power systems through simplification of certain complex components, and for blackbox modelling of off-the-shelf converters.

Several limitations must be addressed before deployment in practical settings. Notably, the proposed model is based on an averaged large-signal representation, which overlooks high-frequency behaviour in predicted signals and neglects non-idealities from diode $D_1$ and switch $S$. Moreover, the dataset relies on an unsaturated controller, limiting the validity of the model by excluding non-linearities in the control. However, our observations indicate that omitting $K_\text{p}$ and $K_\text{i}$ from the feature vector $\textbf{x}_t$ did not significantly impact the performance metrics. This suggests that other controllers could be included into the dataset, as only the control signal remains. While we are confident that this methodology can be applied to other simple dc-dc topologies, such as the buck converter \cite{Ge2023, Qasqai2020}, extending it to more complex topologies is a topic of future work.

Future directions will focus on benchmarking the proposed method against physics-based approaches in terms of computational efficiency to quantify the potential gain in simulation speed. Expanding the framework to include other converter operating modes, such as protective states and fault-tolerant operation, is also a key direction to improve model generalization and extend its applications.

% \section{References}

%For automatic reference with a .bib file use:
\bibliographystyle{IEEEtran}
\bibliography{ref}

\appendix
\section{Appendix A}\label{sec:AppendixA}

We determine $K_\text{p}$ and $K_\text{i}$ using the \textit{algebra on the graph} technique from \cite{Maksimovic2001}. To do so, we first set the cross-over frequency $\omega_{\phi_\text{m}}=2\pi f_{\phi_\text{m}}$ where $f_{\phi_\text{m}}=255$~Hz (more than 200 times smaller than the switching frequency $f_\text{s}$) and the phase-margin $\phi_\text{m}=50\degree$ as a arbitrary compromise between stability margin, overshoot, and settling time. From those desired values, we first calculate the phase and magnitude of $G_\text{c}(s)$ given by
% \begin{align}
    $\angle{G_{\text{c}}(\omega_{\phi_\text{m}})}=-(\angle{}G(\omega_{\phi_\text{m}})+(180\degree-\phi_\text{m}))$ and
    $|G_{\text{c}}(\omega_{\phi_\text{m}})|_\text{dB}=-|G(\omega_{\phi_\text{m}})|_\text{dB}$.
% \end{align}
We then find the ratio $\alpha$ and compute $K_\text{p}^\prime$ of an intermediate controller with $K_\text{i}^\prime=1$ as
\begin{align}
    \alpha=\frac{K_\text{i}}{K_\text{p}} =& \frac{\omega_{\phi_\text{m}}}{\text{tan}(\angle{}G_\text{c}(\omega_{\phi_\text{m}})+90\degree)}\notag\\
    K_\text{p}^{\prime}=&\frac{1}{\alpha}\text{.}\notag
\end{align}
We finally calculate the magnitude of $G_\text{c}^\prime(s)$ to find $K_\text{p}$ and $K_\text{i}$, and verify if a PI-controller with those parameters is feasible:
\begin{align}
    |G_{\text{c}}^\prime(\omega_{\phi_\text{m}})|_\text{dB}=&20\log\left[\frac{\sqrt{(\omega_{\phi_\text{m}}K_\text{p}^\prime)^2+(K_\text{i}^\prime)^2}}{\omega_{\phi_\text{m}}}\right]\notag\\
    K_{i}=&10^{-|G(\omega_{\phi_\text{m}})|_\text{dB}-|G_\text{c}^\prime(\omega_{\phi_\text{m}})|_\text{dB}/20}\notag\\
    K_\text{p}=&\frac{K_\text{i}}{\alpha}\text{.}\notag
\end{align}

\vfill
\end{document}